\documentclass{llncs}

\usepackage[utf8x]{inputenc}
\usepackage{graphicx}
\usepackage{multirow}
\usepackage{subfigure}
\usepackage{color}
\usepackage{enumitem}
\usepackage[linesnumbered,ruled,scleft,nofillcomment]{algorithm2e}
\usepackage{xspace}	
\newcommand{\kdost}{\mathsf{k^2tree}}   %%UNIFICAR AL FINAL
\newcommand{\kdostrees}{\mathsf{k^2trees}}   %%UNIFICAR AL FINAL
\newcommand{\kdosr}{\ensuremath{\mathsf{k^2raster}}\xspace}   %%UNIFICAR AL FINAL
\newcommand{\kdosrp}{\ensuremath{\mathsf{k^2raster'}}\xspace}   %%UNIFICAR AL FINAL
\newcommand{\kdosacc}{\mathsf{k^2acc}}   %%UNIFICAR AL FINAL
\newcommand{\ktrest}{\mathsf{k^3tree}}   %%UNIFICAR AL FINAL
\newcommand{\tkdosr}{\ensuremath{\mathsf{T\!-\!k^2raster}}\xspace}   %%UNIFICAR AL FINAL
\newcommand{\netcdf}{\texttt{NetCDF}\xspace} %%UNIFICAR AL FINAL

\begin{document}
\titlerunning{Towards a compact representation of temporal rasters}
\authorrunning{Cerdeira-Pena et al.}
\authorrunning{A. Cerdeira-Pena, G. de Bernardo, A. Fari\~na, J. Param\'a, and F. Silva-Coira}

\title{Towards a compact representation of temporal rasters
  %
  %\title{Compact Spatio-Temporal Representation for Trips over Networks
   \thanks{
     \footnotesize{
     Funded in part by European Union's Horizon 2020 research and innovation programme
     under the Marie Sklodowska-Curie grant agreement No 690941 (project BIRDS);
	  by Xunta de Galicia/FEDER-UE [CSI: ED431G/01 and GRC: ED431C 2017/58]; 
	  by MINECO-AEI/FEDER-UE [Datos 4.0: TIN2016-78011-C4-1-R; Velocity: TIN2016-77158-C4-3-R; and ETOME-RDFD3: TIN2015-69951-R]; and
	  by MINECO-CDTI/FEDER-UE [INNTERCONECTA: uForest ITC-20161074].	}
   }
}

\author{Ana Cerdeira-Pena \inst{1} \and
	Guillermo de Bernardo \inst{1,2} \and 
    Antonio Fari\~na\inst{1} \and \\
    Jos\'e R. Param\'a\inst{1} \and
    Fernando Silva-Coira\inst{1} %\\
    %\url{\{acerdeira,gdebernardo,fari,parama, fersnado.silva\}@udc.es} 
    }

\institute{Universidade da Coruña, Fac. Inform\'atica, CITIC, Spain \and Enxenio S.L. %\and  University of Chile \and
    }

\maketitle

%%%%%%%%%%%%%%%%%%%%%%%%%%%%%%%%%%%%%%%%%%%%%%%%%%%%%%%%%%%%%%%%%%%%%%%%%%%%%%%%%%%%%%%

%%%%%%%%%%%%%%%%%%%%%%%%%%%%%%%%%%%%%%%%%%%%%%%%%%%%%%%%%%%%%%%%%%%%%%
%%%%%%%%%%%%%%%%%%%%%%%%%%%%%%%%%%%%%%%%%%%%%%%%%%%%%%%%%%%%%%%%%%%%%%
\begin{abstract}
Big research efforts have been devoted to efficiently manage spatio-temporal data. However, most works focused on vectorial data, and much less, on raster data. This work
presents a new representation for raster data that evolve along time  named \texttt{Temporal} $\kdosr$. It  faces  the two main issues that arise when dealing with spatio-temporal data: the space consumption and the query response times. It  
extends a compact data structure for raster data in order to manage time and thus, it is possible to query it directly in compressed form, instead of the classical approach that requires a complete decompression before any manipulation.
In addition, in the same compressed space, the new data structure includes two indexes: a spatial index and an index on the values of the cells, thus becoming a self-index for raster data.

\end{abstract}

%%%%%%%%%%%%%%%%%%%%%%%%%%%%%%%%%%%%%%%%%%%%%%%%%%%%%%%%%%%%%%%%%%%%%%
\section{Introduction}
%%%%%%%%%%%%%%%%%%%%%%%%%%%%%%%%%%%%%%%%%%%%%%%%%%%%%%%%%%%%%%%%%%%%%%

Spatial data can be represented using either a raster or a vector data model \cite{Couclelis92}. Basically, vector models represent the space using points and lines connecting those points. They  are used mainly to represent man-made features. Raster models  represent the space as a tessellation of disjoint fixed size tiles (usually squares), each one storing a value. They are traditionally used  in engineering, modeling, and representations of real-word elements that were not made by men, such as pollution levels, atmospheric and vapor pressure, temperature, precipitations, wind speed, land elevation, satellite imagery,  etc. 

Temporal evolution of vectorial data has been extensively studied, with a large number of data structures to index and/or store spatio-temporal data. Examples are  the 3DR-tree \cite{Vazirgiannis1998}, HR-tree \cite{NascimentoS98}, the MVR-tree \cite{PapadiasT01}, or PIST \cite{Botea2008}. 

In \cite{mennis} the classical Map Algebra of Tomlin for managing raster data is extended to manage raster data with a temporal evolution. The conceptual solution is simple, instead of considering a matrix, it considers a cube, where each slice of the temporal dimension is the  raster corresponding to one time instant. 

Most real systems capable of managing raster data, like Rasdaman, Grass, or even R are also capable of managing time-series of raster data. These systems, as well as raster representation formats such as NetCDF (standard format of the OGC\footnote{\url{http://www.opengeospatial.org/standards/netcdf}}) and GeoTiff, rely on classic compression methods such as run length encoding, LZW, or Deflate to reduce the size of the data. The use of these compression methods poses an important drawback to access a given datum or portion of the data, since the dataset must be decompressed from the beginning.

Compact data structures \cite{j-ssds-89,Nav16} are capable of storing data in compressed form and enable us to access a given datum without the need for decompressing from the beginning. In most cases, compact data structures are equipped with an index that provides fast access to data. There are several compact data structures designed to store raster data \cite{k2ones,LadraGonzalez17}. In this work, we extend one of those compact data structures, the $\kdosr$ \cite{LadraGonzalez17}, to support representing time-series of rasters.

%%%%%%%%%%%%%%%%%%%%%%%%%%%%%%%%%%%%%%%%%%%%%%%%%%%%%%%%%%%%%%%%%%%%%%%%%%%%%%%%%%

%%%%%%%%%%%%%%%%%%%%%%%%%%%%%%%%%%%%%%%%%%%%%%%%%%%%%%%%%%%%%%%%%%%%%%
\section{Related work} \label{sec:prevwork}
%%%%%%%%%%%%%%%%%%%%%%%%%%%%%%%%%%%%%%%%%%%%%%%%%%%%%%%%%%%%%%%%%%%%%%

In this section, we first revise the $\kdost$, a compact data structure that can be used to represent binary matrices. Then, we also present several compact data structures for representing raster data containing integers in the cells. We pay special attention to discuss one of them, the $\kdosr$, which is the base of our proposal \texttt{Temporal} $\kdosr$ ($\tkdosr$).

\vspace{-4mm}
\subsubsection{$\kdost$:} 
%\paragraph{\textbf{{$\kdost$:} }}
The $\kdost$ \cite{ktree} was initially designed to %compactly 
represent web graphs,  but it also allows to represent binary matrices, that is,  rasters where the cells  contain only a bit value. It is conceptually a non-balanced $k^2$-ary tree built from the binary matrix by recursively dividing it into
$k^2$ submatrices of the same size. First,  the original matrix is divided into
$k^2$ submatrices of size $n^2/k^2$, being $n$$\times$$n$ the size of the matrix.  Each  submatrix  generates a child of the root whose value is $1$  if it contains at least one $1$, and $0$ otherwise. The subdivision continues recursively for each node representing a submatrix that has at least one  $1$, until the submatrix is full of 0s, or until the process reaches the cells of the original matrix (i.e., submatrices of size 1$\times$1). %Figure~\ref{fig:example} shows an example of this subdivision (left) and the resulting conceptual $k^2$-ary tree (right up) for $k=2$. 

The $\kdost$ is compactly stored using just two bitmaps $T$ and $L$. 
%(see Figure \ref{fig:example}).
\textit{T}  stores all the bits of the conceptual $\kdost$, except the last level,  following a level-wise traversal: first the bit values of the children of the root, then those in the second level, and so on. $L$ stores the last level of the tree. 

It is possible to obtain any cell, row, column, or window of the matrix very efficiently, by running $rank$  and $select$ operations \cite{j-ssds-89} over the bitmaps $T$ and $L$.

% \begin{figure}[t]
% 	\begin{center}	
% 		\includegraphics[scale=0.15]{figures/k2-base.pdf} 	
% 	\end{center}
%     \vspace{-3mm}
% 	\caption{Example of binary raster (left), the $k^2$-tree conceptual representation (top right), and  the compact representation (bottom right), where $k=2$.}
%     %\vspace{-3mm}
% 	\label{fig:example}
% \end{figure}

\vspace{-4mm}
\subsubsection{$\ktrest$:}
The $\ktrest$ \cite{k2ones} is obtained by simply adding a third dimension to the $\kdost$, and thus, it conceptually represents a binary cube.  This can be trivially done by using the same space partitioning and representation techniques from the $\kdost$, yet applied to cubes rather than to matrices.

Thus, each 1 in the binary cube  represents a tuple $\langle x, y,z \rangle$, where %in our case, 
$(x,y)$ are the coordinates of the cell of the raster and $z$ is the value stored in that cell.
%It is also possible to  obtain the value of a cell, a cube, or slices of the cube by fixing query values when required.

\vspace{-4mm}
\subsubsection{$\kdosacc$:}
The $\kdosacc$ \cite{k2ones} representation for a raster dataset is  composed by as many $\kdostrees$ as different  values can be found in the raster.
Given  $t$ different values in the raster: $v_1 < v_2 < \dots < v_t$,
$\kdosacc$ contains  $K_1,K_2,\dots,K_t$ $\kdostrees$, where each $K_i$ has a value 1  in those
cells whose value is $v \le v_i$. 

%This representation can return the value at a given cell by binary searching the collection
%of $\kdostrees$, and can also efficiently return those cells within a given range of values by accessing only the $\kdostrees$ of the two ends of the range.

\vspace{-4mm}
\subsubsection{2D-1D mapping:}
In \cite{Pinto17}, it is presented a method that uses a space-filling curve to reduce the raster matrix to an array, and the use of one dimensional index (for example a B-tree) over that array to access the data.

\vspace{-4mm}
\subsubsection{$\kdosr$:}
%We are going to describe the compact data structure $\kdosr$, since it is the  base raster representation for our method.
$\kdosr$ has proven to be superior in both space and query time \cite{Pinto17,LadraGonzalez17} to all the other compact data structures for storing rasters. In \cite{LadraGonzalez17}, it was also compared with NetCDF. 
 %The results showed a space consumption slightly worse than that of the compressed version of NetCDF (that uses Deflate). Yet, query times were noticeably much better.  
It drew slightly worse space needs than the compressed version (that uses Deflate) of NetCDF, but queries performed noticeably faster.

%$\kdosr$ is able to represent a matrix having integer values in the cells, and efficiently supports queries directly over the compressed data. It 
$\kdosr$ 
is based in the same partitioning method of the $\kdost$, that is,
it recursively subdivides the matrix into $k^2$ submatrices and builds a conceptual tree representing these subdivisions. Now, in each node, instead of having a single bit, it contains  the minimum and maximum values of the corresponding submatrix. The subdivision stops when the minimum and maximum values of the submatrix are equal, or when the process reaches submatrices of size 1$\times$1. 
Again the conceptual tree is compactly represented using, in addition to binary bitmaps, efficient encoding schemes for integer sequences.

\begin{figure}[th]
  \begin{center}
  \begin{tabular}{c}	
  \includegraphics[width=0.9\textwidth]{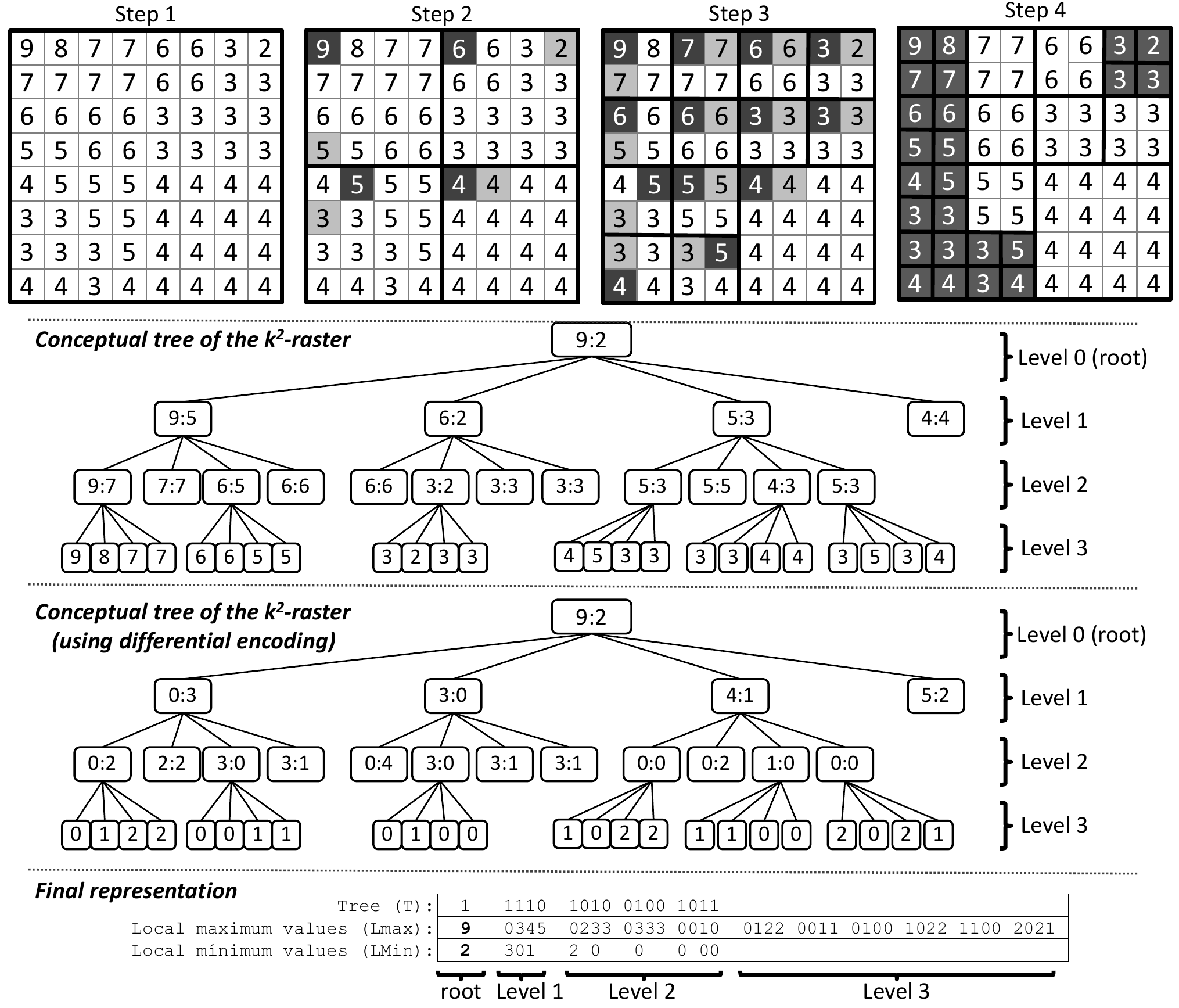}\\
  \end{tabular}
  \end{center}
  \vspace{-5mm}
  \caption{Example (using $k=2$) of integer raster matrix (top), conceptual tree of the $\kdosr$, conceptual tree with differential encoding, and final representation of the raster matrix. $Lmax$ and $Lmin$ contain the maximum and minimum values of each node following a level-wise order and using differential encoding.
}
\label{fig:k2raster}
\end{figure}

More in detail, let $n$$\times$$n$ be the size of the input matrix.  The process begins by obtaining the minimum and maximum values of the matrix. If these values are different, they are stored in the root of the tree, and the matrix is  divided into $k^2$ submatrices of size $n^2/k^2$. Each submatrix produces  a child node of the root storing its minimum and maximum values. If these values are the same, that node becomes a leaf, and the  corresponding submatrix is not  subdivided anymore. Otherwise, this procedure continues recursively until the maximum and minimum values are the same, or the process reaches a 1$\times$1 submatrix.

%\OJO{ polo que vexo despois, en T, non se mete o 1 do root. habera que cambiar isto na Fig.\ref{fig:k2raster}, ou se non, adaptar o algoritmo getCell}

Figure \ref{fig:k2raster} shows an example of the recursive subdivision (top) and how the conceptual tree is built (centre-top), where the minimum and maximum values of each submatrix are stored at each node. The root node corresponds to the original raster matrix, nodes at level 1 correspond to submatrices of size 4$\times$4, and so on. The last level of the tree corresponds to cells of the original matrix. Note, for instance, that all the values of the bottom-right 4$\times$4 submatrix are equal; thus, its minimum and maximum values are equal, and it is not further subdivided. This is the reason why the last child of the root node has no children. 

The compact representation includes two main parts. The first one represents the topology of the tree ($T$) and the second one stores the maximum/minimum values at the nodes ($Lmin/Lmax$). 
The topology is represented as in the $\kdost$, except that the last level ($L$) is not needed. The maximum/minimum values are differentially encoded with respect to the values stored at the parent node.
Again, these values are stored as arrays following the same method of the $\kdost$, that is, following the same level-wise order of the conceptual tree. By using differential encoding, the numbers become smaller. \emph{Directly Addressable Codes} (DACs) \cite{BLN13} take advantage of this, and at the same time, provide direct access. % to any given value without decompressing the rest.
The last two steps to create the final representation of the example matrix are also illustrated in Figure \ref{fig:k2raster}. 
%In the center-bottom part of the figure we show the tree with the differences for both the maximum and minimum values, whereas the data structures that compose the final representation of the $k^2$-raster are shown at the bottom part. 
In the center-bottom and bottom parts, we respectively show the tree with the differences for both the maximum and minimum values, and the data structures that compose the final representation of the $\kdosr$. 
Therefore, the original  raster matrix is compactly stored using just a bitmap $T$, which represents the tree topology, and a pair of integer arrays ($Lmax$ and $Lmin$), which contain the minimum and maximum values stored at the tree. Note that when the raster matrix contains uniform areas, with large areas of equal or similar values, this information can be  stored very compactly using differential encoding and DACs.

The maximum/minimum values provide indexation of the stored values, this technique is usually known as lightweight indexation. It is possible to query the structure only decompressing the affected areas. Queries can be efficiently computed navigating the conceptual tree by running $rank$ and $select$ operations on $T$ and, in parallel, accessing the arrays $Lmax$ and $Lmin$. 

%\OJO{Esto se pude quitar tranquilamente}
%There are two variants of the basic $\kdosr$ \cite{LadraGonzalez17} that improve its space/time trade-off. 

%The first one uses different $k$ values for each level. In practice, it uses only two values $v_1$ and $v_2$: $v_1$ is used from the root until a given level, and $v_2$ from there on. This method typically requires slightly more space, but answer queries faster. The second alternative is to use a vocabulary of leaf submatrices. If a submatrix occurs frequently as a leaf in  several branches of the tree, it is stored in a vocabulary, and its occurrences  are replaced by a pointer to the corresponding vocabulary entry. This variant improves both space and time efficiency.

%%%%%%%%%%%%%%%%%%%%%%%%%%%%%%%%%%%%%%%%%%%%%%%%%%%%%%%%%%%%%%%%%%%%%%%%%%%%%%%
\section{$\tkdosr$: A temporal representation for raster data}
%%%%%%%%%%%%%%%%%%%%%%%%%%%%%%%%%%%%%%%%%%%%%%%%%%%%%%%%%%%%%%%%%%%%%%%%%%%%%%
Let $\cal{M}$ be a raster matrix of size $n$$\times$$n$ that evolves along time with a timeline of size $\tau$ time instants. We can define $\cal{M}$$= \langle M_1, M_2, \dots, M_{\tau} \rangle$ as the sequence of raster matrices $M_i$ of size $n$$\times$$n$ for each time instant $i \in [1,\tau]$. 

A rather straightforward baseline representation for the temporal raster matrix $\cal{M}$ can be obtained by simply representing each raster matrix $M_i$ in a compact way with a $\kdosr$. In this section we use a different approach. The idea is to use sampling at regular intervals of size $t_{\delta}$. That is,  we represent with a $\kdosr$ all the raster matrices $M_s, s=1+i~t_{\delta},~i\in [0,(\tau-1) /t_{\delta}]$. We will refer to those $M_i$ rasters as {\em snapshots of $\cal{M}$ at time $i$}. The $t_{\delta} -1$ raster matrices $M_t, t \in [s+1, s+t_{\delta} -1]$ that follow a snapshot $M_s$ are encoded using $M_s$ as a reference. The idea is to create  a modified $\kdosrp$ to represent $M_t$ where, at each step of the construction process, the values in the submatrices are encoded as differences with respect to the corresponding submatrices in $M_s$ rather than as  differences with respect to the parent node as usual in a regular $\kdosr$. 

With this modification, we still expect to encode small gaps for the maximum and minimum values in each node of the conceptual tree of $M_t$. Yet, in addition, when a submatrix in $M_t$ is identical to the same submatrix in $M_s$, or when all the values in both submatrices differ only in a unique gap value $\alpha$, we can stop the recursive splitting process and simply have to keep a reference to the corresponding submatrix of $M_s$ and the gap $\alpha$ (when they are identical, we simply set $\alpha=0$). In practice, keeping that reference is rather cheap as we only have to mark, in the conceptual tree of $M_t$, that the subtree rooted at a given node $p$ has the same structure of the one from the conceptual tree of $M_s$. 
For such purpose, in the final representation of $\kdosrp$, we include a new bitmap $eqB$, aligned to the zeroes in $T$. That is, if we have $T[i]=0$ (node with no children), we set $eqB[rank_0(T,i)] \leftarrow 1 $,\footnote{From now on, asume $rank_b(B,i)$ returns the number of bits set to $b$ in $B[0,i-1]$, and $rank_b(B,0)=0$. Note that the first index of $T$, $eqB$, $Lmax$, and $Lmin$ is $0$. } and set $Lmax[i] \leftarrow \alpha$. Also, if we have $T[i]=0$, we also can set $eqB[rank_0(T,i)] \leftarrow 0 $ and $Lmax[i]\leftarrow \beta$ (where $\beta$ is the gap between the maximum values of both submatrices) to handle the case in which the maximum and minimum values in the corresponding submatrix are identical (as in a regular $\kdosr$). 

The overall construction process of the $\kdosrp$ for the matrix $M_t$ related to the snapshot $M_s$ can be summarized as follows. 
At each step of the recursive process, we consider a submatrix of $M_t$ and the related submatrix in $M_s$. Let the corresponding maximum and minimum values of the submatrix of $M_t$ be $max_t$ and $min_t$, and those of $M_s$ be $max_s$ and $min_s$ respectively. Therefore:
\begin{itemize}
	\item If $max_t$ and $min_t$ are identical (or if we reach a 1$\times$1 submatrix), the recursive process stops. Being $z_t$ the position in the final bitmap $T$, we set $T[z_t]\leftarrow 0$, $eqB[rank_0[T,z_t]]\leftarrow 0$, and $Lmax[z_t] \leftarrow (max_t -max_s)$.\footnote{Since in $\kdosrp$ we have to deal both with positive and negative values, we actually apply a {\em zig-zag} encoding for the gaps $(max_t -max_s)$.} 
	\item If all the values in $M_s$ and $M_t$ differ only in a unique value $\alpha$ (or if they are identical, hence $\alpha=0$), we set $T[z_t]\leftarrow 0$, $eqB[rank_0[T,z_t]]\leftarrow 1$, and $Lmax[z_t] \leftarrow (max_t -max_s)$.
	\item Otherwise, we split the submatrix $M_t$ into $k^2$ parts and continue recursively. We set $T[z_t]\leftarrow 1$, and, as in the regular $\kdosr$,   $Lmax[z_t]\leftarrow (max_t -max_s)$, and $Lmin[rank_1(z_t)]\leftarrow (min_t - min_s)$.  
\end{itemize}
	
\begin{figure}[ht]
	\vspace{-0.4cm}
	\begin{center}
		{\includegraphics[width=1.00\textwidth]{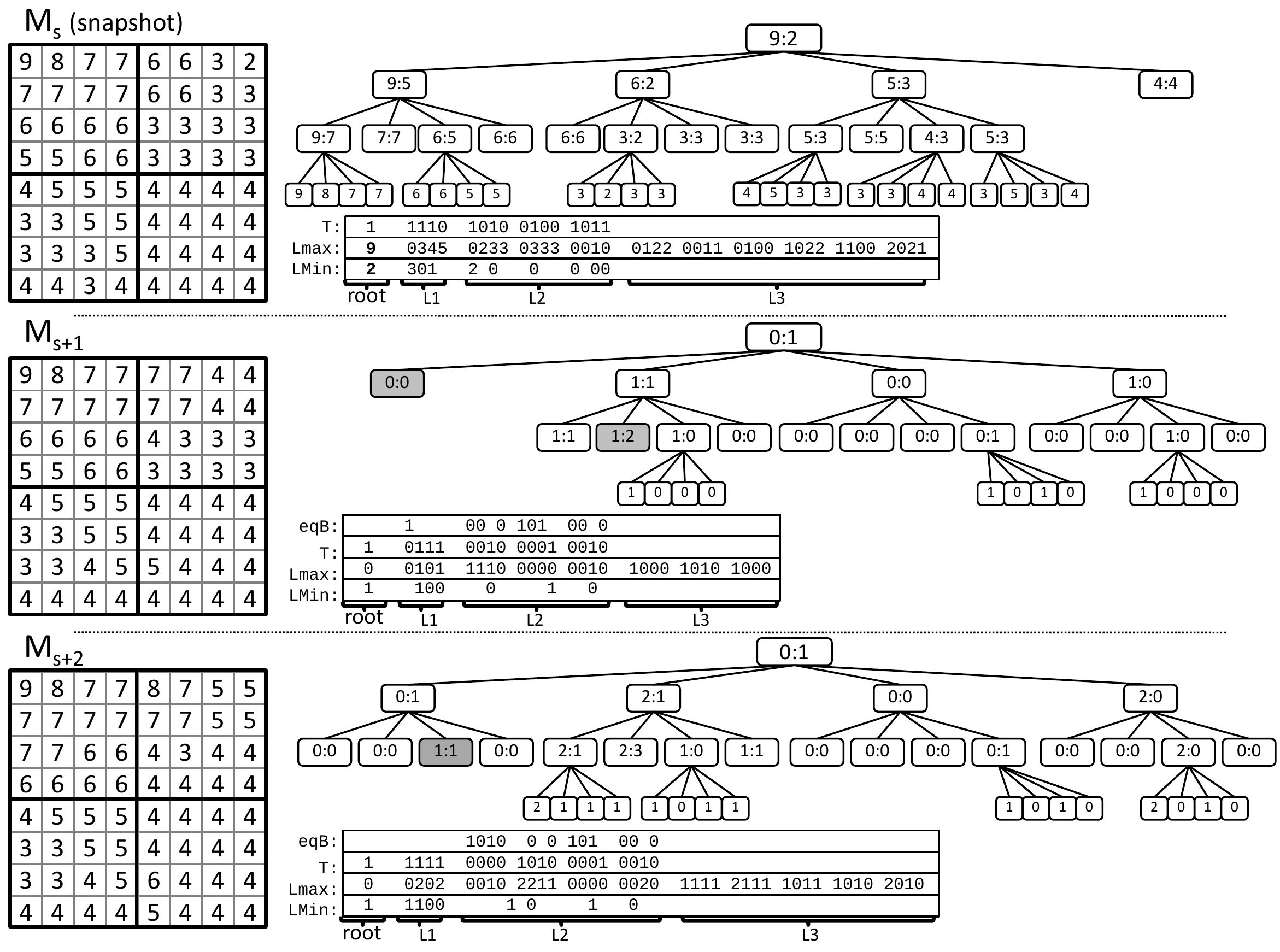}}
	\end{center}
	\vspace{-0.4cm}
	\caption{Structures involved in the creation of a $\tkdosr$ considering $\tau=3$.}
	\label{fig:tkdosr}
	\vspace{-0.5cm}
\end{figure}
	
Figure~\ref{fig:tkdosr} includes an example of the structures involved in the construction of a $\tkdosr$ over a temporal raster of size 8$\times$8, with $\tau=3$. The raster matrix corresponding to the first time instant becomes a {\em snapshot} that is represented exactly as the $\kdosr$ in Figure~\ref{fig:k2raster}. The remaining raster matrices $M_{s+1}$ and $M_{s+2}$ are represented with two $\kdosrp$ that are built taking $M_s$ as a reference. We have highlighted some particular nodes in the differential conceptual trees corresponding to $M_{s+1}$ and $M_{s+2}$. 
{\em (i)} the shaded node labeled $\langle 0 \colon 0 \rangle$ in $M_{s+1}$ indicates that the first 4$\times$4 submatrix of both $M_s$ and $M_{s+1}$ are identical. Therefore, node  $\langle 0 \colon 0 \rangle$ has no children, and we set: $T[2]\leftarrow 0$, $eqB[1]\leftarrow 1$, and $Lmax[2]\leftarrow 0$.
{\em (ii)} the shaded node labeled $\langle 1 \colon 1 \rangle$ in $M_{s+2}$ illustrates the case in which all the values of a given submatrix are increased by $\alpha \leftarrow 1$. In this case values $\langle 6,6,5,5 \rangle$ in $M_s$ become $\langle 7,7,6,6 \rangle$ in $M_{s+2}$. Again, the recursive traversal stops at that node, and we set: $T[8]\leftarrow 0$, $eqB[3]\leftarrow 1$, and $Lmax[8]\leftarrow 1$ (values are increased by $1$).
{\em (iii)} the shaded node labeled $\langle 1 \colon 2 \rangle$ in $M_{s+1}$ corresponds to the node labeled $\langle 3 \colon 2 \rangle$ in $M_s$. In this case, when we sum the maximum and minimum values of both nodes we obtain that that node in $M_{s+1}$ has the same maximum and minimum values (set to $4$). Consequently the recursive process stops again. In this case, we set $T[7]\leftarrow 0$, $eqB[3]\leftarrow 0$, and $Lmax[7]\leftarrow 1$.

%%%%%%%%%%%%%%%%%%%%%%%%%%%%%%%%%%%%%%%%%%%%%%%%%%%%%%%%%%%%%%%%%%%%%%%%%%%%%%%%%%
\section{Querying temporal raster data} \label{sec:tcsaqueries}
%%%%%%%%%%%%%%%%%%%%%%%%%%%%%%%%%%%%%%%%%%%%%%%%%%%%%%%%%%%%%%%%%%%%%%%%%%%%%%%%%%

In this section, we show  two basic queries over  $\tkdosr$. 
% \begin{itemize}
% \item $v\leftarrow getCellValue(r,c,t)$. 
% \item $\langle[r_i,c_i]\rangle this query,r_1,r_2,c_1,c_2,t)$
% \end{itemize}

% \OJOFARI{Quizais poidamos decir algo como isto ao final: 
% Note that those queries could be also defined for a time interval $[t_1,t_2]$. In that case, we would have to traverse the corresponding incremental $\kdosr$ corresponding to the times $t_i \in [t_1,t_2]$ to compute the {\em union} of those trees. We could benefit of stopping the traversal when we reach cells already included in the resulting set of cells. ¿non si '???
% }

\vspace{-4mm}
\subsubsection*{Obtaining a cell value in a time instant:}
This query retrieves the value of a cell $(r,c)$ of the raster at time instant $t$: $v\leftarrow getCellValue(r,c,t)$. 
For solving this query, there are two cases: if $t$ is represented by a snapshot, then the algorithm  to obtain a cell in the regular $\kdosr$ is used, otherwise, a synchronized top-down traversal of the trees representing that time instant ($M_{t}$) and the closest previous snapshot ($M_{s}$) is required.

Focusing on the second case, the synchronized traversal inspects the two nodes at each level corresponding to the submatrix that contains the queried cell. The problem is that due to parts of $M_{t}$ or $M_{s}$ having the same value, the shape of the trees representing them can be different. Therefore, it is possible that one of the two traversals reaches a leaf, whereas the other does not. In such a case, the traversal that did not reach a leaf, continues, but the process must remember the value in the reached leaf, since that is the value that will be  added or subtracted to the value found when the continued traversal reaches a leaf.

Indeed, we have three cases: (a) the processed submatrix of $M_t$ is uniform, (b) the original submatrix of $M_s$  is uniform and, (c) the processed submatrix after applying the differences  with the snapshot has the same value in all cells. 

% To access the value of a given position, the $k^2$-$raster$ performs a top-down
% traversal of the tree, guided by the quadrant where the cell is located at each level. While going down in the tree,
% we must decode the maximum values stored for the traversed nodes and subtracting them from the maximum value stored at the root node.

\tolerance 10000 \pretolerance 10000
Algorithm \ref{alg:getCell} shows the pseudocode of this case. To obtain the value stored at cell $(r,c)$ of the raster matrix $M_{t}$, it is invoked as {\bf getCell}$(n,r,c,1,1,Lmax_s[0],Lmax_t[0])$,  assuming that the  cell at position (0,0) of the raster is that in the upper-left corner. 

\tolerance 500 \pretolerance 500

 $z_s$ is used to store the current position in the bitmap $T$ of $M_{s}$ ($T_{s}$) during the downward traversal at any given step of the algorithm, similarly,  $z_t$ is the position in $T$ of $M_{t}$ ($T_{t}$). When  $z_s$ ($z_t$) has a $-1$ value, it means that the traversal reached a leaf and, in $maxval_s$ ($maxval_t$) the algorithm keeps the maximum value stored at that leaf node. %The initial $-2$ is an artifact because $T$ does not represent the root node. It is assumed that $rank(T,-2)=0$.
Note that, $T_s$, $T_t$, $Lmax_s$, $Lmax_t$, and $k$ are global variables.

% Algorithm \ref{alg:getCell} shows the pseudocode for this query. To obtain the value at
% position $(x,y)$ of the raster matrix, that is $M_{xy}$, it is invoked as {\bf getCell}$(n,x,y,-1,rMax)$.
% It consists in a recursive procedure whose parameters are: current submatrix size,
% row of interest in the current submatrix, column of interest in the current submatrix, and
%  the position in $T$ of the node to process (the
% initial $-1$ is an artifact because $T$ does not represent the root node).
% Values $T$, $Lmax$, and $k$ are global. It is assumed that $rank(T,-1)=0$ and $rMax$ is the
% maximum value of the raster matrix, which is not represented in $Lmax$.

\begin{algorithm}[t]
\scriptsize
\SetAlgoNoLine
\caption{{\bf getCell}$(n,r,c,z_s,z_t, maxval_s,maxval_t)$ returns the value at cell $(r,c)$}\label{alg:getCell}

\If{$z_s\neq -1$}{
 $z_s \leftarrow (rank_1(T_s,z_s)-1) \cdot k^2+1$ \\
 $z_s \leftarrow z_s + \lfloor r/(n/k) \rfloor\cdot k + \lfloor c/(n/k) \rfloor$ +1\\
 $val_s \leftarrow Lmax_s[z_s-1]$\\
 $maxval_s \leftarrow maxval_s-val_s$\\
}

\If{$z_t\neq -1$}{
 $z_t \leftarrow (rank_1(T_t,z_t)-1) \cdot k^2$+1 \\
 $z_t \leftarrow z_t + \lfloor r/(n/k) \rfloor\cdot k + \lfloor c/(n/k) \rfloor$ +1\\
 $maxval_t \leftarrow Lmax_t[z_t-1])$\\
}

\If(\tcc*[h]{Both leafs}){$(z_s > |T_s| ~\mathbf{or}~z_s=-1 ~\mathbf{or}~ T_s[z_s]=0)~\mathbf{and}~(z_t > |T_t| ~\mathbf{or}~z_t=-1 ~\mathbf{or}~ T_t[z_t]=0)$}{
\Return $maxval_s+ZigZag\_Decoded(maxval_t)$\\
}
\ElseIf(~\tcc*[h]{Leaf in Snapshot}){$z_s > |T_s| ~\mathbf{or}~z_s=-1 ~\mathbf{or}~ T_s[z_s]=0$}
{$z_s \leftarrow -1$\\
\Return {\bf getCell}$(n/k,r\,\textrm{mod}\,(n/k),c\,\textrm{mod}\,(n/k),z_s,z_t, maxval_s,maxval_t)$}
\ElseIf(~\tcc*[h]{Leaf in time instant}){$z_t > |T_t| ~\mathbf{or}~z_t=-1 ~\mathbf{or}~ T_t[z_t]=0$}
{\If{$z_t\neq -1 ~\mathbf{and}~ T_t[z_t]=0$}{$eq \leftarrow eqB[rank_0(T_t,z_t)]$\\
                   \lIf{$eq=1$}
                     {$z_t\leftarrow -1$ }
                    \lElse{\Return $maxval_s+ZigZag\_Decoded(maxval_t)$   
                    }  
                   }
\Return {\bf getCell}$(n/k,r\,\textrm{mod}\,(n/k),c\,\textrm{mod}\,(n/k),z_s,z_t, maxval_s,maxval_t)$\\} 
\Else(~\tcc*[h]{Both internal nodes}) 
{
\Return {\bf getCell}$(n/k,r\,\textrm{mod}\,(n/k),c\,\textrm{mod}\,(n/k),z_s,z_t, maxval_s,maxval_t)$\\
}

\end{algorithm}
%\vspace{-5mm}

In lines 1-11, the algorithm obtains the child of the processed node that contains the queried cell, provided that in a previous step, the algorithm did not reach a leaf node (signaled with $z_s$/$z_t$ set to $-1$). In $maxval_s$ ($maxval_t$), the algorithm stores the maximum value stored in that node.

If the condition in line 12 is true,  the algorithm has reached a leaf in both trees, and thus the values stored in $maxval_s$  and $maxval_t$ are added/subtracted  to obtain the final result.
If the condition of line 15 is true, the algorithm reaches a leaf in the snapshot. This is signaled by setting $z_s$ to $-1$ and then a recursive call continues the process.

The {\em If} in line 19 treats the case of reaching a leaf in $M_t$. If the condition of line 20 is true,  the algorithm uses bitmap $eqB$ to check if the uniformity is in the original $M_t$ submatrix or if it is in the submatrix resulting from applying the differences between the corresponding submatrix in $M_s$ and $M_t$. A $1$ in $eqB$ implies the latter case, and this is solved by setting $z_t$ to $-1$ and performing a recursive call. A $0$ means that the treated original submatrix of $M_t$ has the same value in all cells, and that value can be obtained adding/subtracting the values stored in $maxval_s$  and $maxval_t$, since the unique value in the  submatrix of $M_t$ is encoded as a difference with respect to the maximum value of the same submatrix of $M_s$, and thus the traversal ends. 

The last case is that the treated nodes are not leaves, that simply requires a recursive call.

\subsubsection*{Retrieving cells with range of values in a time instant:}

$\langle[r_i,c_i]\rangle \leftarrow getCells(v_b,v_e,r_1,r_2,c_1,c_2,t)$  obtains from the raster of the time instant $t$, the positions of all cells within a region $[r_1,r_2]\times [c_1,c_2]$ containing values in the range $[v_b,v_e]$.

Again, if $t$ is represented with a snapshot, the query is solved with the normal algorithm of the $\kdosr$. Otherwise,  as in the previous query, the search involves a synchronized top-down traversal of both trees. This time  requires two main changes: (i) the traversal probably requires following several branches of both trees, since the queried region can overlap the submatrices corresponding to several nodes of the tree, (ii) at each level, the algorithm has to  check whether the maximum and minimum values in those submatrices are compatible with the queried range, discarding those that fall outside the range of values sought.

%%%%%%%%%%%%%%%%%%%%%%%%%%%%%%%%%%%%%%%%%%%%%%%%%%%%%%%%%%%%%%%%
\section{Experimental evaluation} \label{sec:experiments}
%%%%%%%%%%%%%%%%%%%%%%%%%%%%%%%%%%%%%%%%%%%%%%%%%%%%%%%%%%%%%%%%
In this section we provide experimental results to show how $\tkdosr$ handles a dataset of raster data that evolve along time. We discuss both the space requirements of our representation and its performance at query time. 

We used several synthetic and real datasets to test our representation, in order to show its capabilities. All the datasets are obtained from the TerraClimate collection~\cite{TerraClimate}, that contains high-resolution time series for different variables, including temperature, precipitations, wind speed, vapor pressure, etc. All the variables in this collection are taken in monthly snapshots, from 1958 to 2017. Each snapshot is a 4320$\times$8640 grid storing values with  $1/24^\circ$ spatial resolution.
From this collection we use data from two variables: TMAX (maximum temperature) is used to build two synthetic datasets, and VAP (vapor pressure) is compressed directly using our representation. Variable TMAX is a bad scenario for our approach, since most of the cells change their value between two snapshots. In this kind of dataset, our $\tkdosr$ would not be able to obtain good compression. Hence, we use TMAX to generate two synthetic datasets that simulate a slow, and approximately constant, change rate, between two real snapshots. We took the snapshots for January and February 2017 and built two synthetic datasets called T\_100 and T\_1000, simulating 100 and 1000 intermediate steps between both snapshots; however, note that to make comparisons easier we only take the first 100 time steps in both datasets. We also use a real dataset, VAP, that contains all the monthly snapshots of the variable VAP from 1998 to 2017. 
Note that, although we choose a rather small number of time instants in our experiments, the performance of our proposal is not affected by this value: it scales linearly in space with the number of time instants, and query times should be unaffected as long as the change rate is similar.

% \renewcommand{\arraystretch}{1.5}
% \begin{table}[th]
%     \centering
%     \setlength\tabcolsep{6pt}
%      %\scalebox{0.95}{
%      	\scriptsize  
% %    \begin{tabular}{|l|c|c|c|}
% %    \hline
% %    Dataset	&	Type	&	Size	& Time instants	\\
% %    \hline
% %    \texttt{T\_100}	&	Synthetic	&	4320$\times$8640	& 100	\\
% %    \texttt{T\_1000}	&	Synthetic	&	4320$\times$8640	& 100	\\
% %    \texttt{VAP}		&	Real		&	4320$\times$8640	& 240	\\
% % 	\hline
% %    \end{tabular}
% 	\begin{tabular}{|l|c|c|}
%    \hline
%    Dataset	&		Size	& Time instants	\\
%    \hline
%    \texttt{T\_100}	&		4320$\times$8640	& 100	\\
%    \texttt{T\_1000}	&		4320$\times$8640	& 100	\\
%    \texttt{VAP}		&		4320$\times$8640	& 240	\\
% 	\hline
%    \end{tabular}
   
%    %\normalsize
% % }
% \caption{Datasets used in our experiments}\label{table:datasets}

% \end{table}

We compared our representation with two baseline implementations. The first, called $\kdosr$\footnote{https://gitlab.lbd.org.es/fsilva/k2-raster} is a representation that stores just a full snapshot for each time instant, without trying to take advantage of similarities between close time instants. The second baseline implementation, \netcdf, stores the different raster datasets in NetCDF format, using straightforward algorithms on top of the NetCDF library\footnote{https://www.unidata.ucar.edu/software/netcdf/} (v.4.6.1) to implement the query operations. Note that NetCDF is a classical representation designed mainly to provide compression, through the use of Deflate compression over the data. Therefore, it is not designed to efficiently answer indexed queries.

We tested cell value queries (\textit{getCellValue}) and range queries (\textit{getCells}). We generated sets of 1000 random queries for each query type and configuration: 1000 random cell value queries per dataset, and sets of 1000 random range queries for different spatial window sizes (ranging from 4$\times$4 windows to the whole matrix), and different ranges of values (considering cells with 1 to 4 possible values). To achieve accurate results, when the total query time for a query set was too small, we repeated the full query set a suitable number of times (in practice, 100 or 1000 times) and measured the average time per query. 

All tests were run on an Intel (R) Core TM i7-3820 CPU @ 3.60GHz (4 cores)
with 10MB of cache and 64GB of RAM, over Ubuntu 12.04.5 LTS with kernel
3.2.0-126 (64 bits). The code is compiled using gcc 4.7 with -O9 optimizations.

\renewcommand{\arraystretch}{1.3}

\begin{table}[th]
\vspace{-5mm}
    \centering
    \setlength\tabcolsep{3pt}
     %\scalebox{0.95}{
     	{\scriptsize  
   \begin{tabular}{l|cccccc|c|cccc|}
         \cline{2-12}
              & \multicolumn{6}{|c|}{\textbf{\tkdosr} (varying $t_\delta$)} & \multirow{2}{*}{\textbf{\kdosr}} & \multicolumn{4}{|c|}{\textbf{\netcdf} (varying deflate level)}\\
              & \multicolumn{1}{|c}{\textbf{4}} & \multicolumn{1}{c}{\textbf{6}} & \multicolumn{1}{c}{\textbf{8}} & \multicolumn{1}{c}{\textbf{10}} & \multicolumn{1}{c}{\textbf{20}} & \multicolumn{1}{c|}{\textbf{50}} & & \multicolumn{1}{|c}{\textbf{0}} & \multicolumn{1}{c}{\textbf{2}} & \multicolumn{1}{c}{\textbf{5}} & \multicolumn{1}{c|}{\textbf{9}}\\
         \hline

\multicolumn{1}{|l|}{\texttt{T\_100}} & 398.2 & 407.0 & 429.6 & 456.7 & 584.4 & 820.8 & 769.3 & 14241.3 & 615.3 & 539.5 & 528.0 \\
        \hline
\multicolumn{1}{|l|}{\texttt{T\_1000}} & 170.4 & 152.5 & 151.2 & 154.6 & 196.2 & 304.6 & 496.6 & 14241.3 & 435.0 & 344.7 & 323.6  \\
        \hline

\end{tabular}
}
%\normalsize
% }
    \caption{Space requirements (in MB) of \tkdosr, \kdosr and \netcdf over synthetic datasets.}\label{table:syntheticsizes}
    \vspace{-8mm}
\end{table}

Table~\ref{table:syntheticsizes} displays the space requirements for the datasets T\_100 and T\_1000 in all the representations. We tested our \tkdosr with several sampling intervals $t_\delta$, and also show the results for $\netcdf$ using several deflate levels, from level 0 (no compression) to level 9. Our representation achieves the best compression results in both datasets, especially in T\_1000, as expected, due to the slower change rate. In T\_100, our approach achieves the best results for $t_\delta = 4$, since as the number of changes increases our differential approach becomes much less efficient. In T\_1000, the best results are also obtained for a relatively small $t_\delta$ (6-8), but our proposal is still smaller than $\kdosr$ for larger $t_\delta$. $\netcdf$ is only competitive when compression is applied, otherwise it requires roughly 20 times the space of our representations. In both datasets, $\netcdf$ with compression enabled becomes smaller than the $\kdosr$ representation, but $\tkdosr$ is able to obtain even smaller sizes.

\begin{figure}[htbp]
\vspace{-8mm}
  \includegraphics[width=\textwidth]{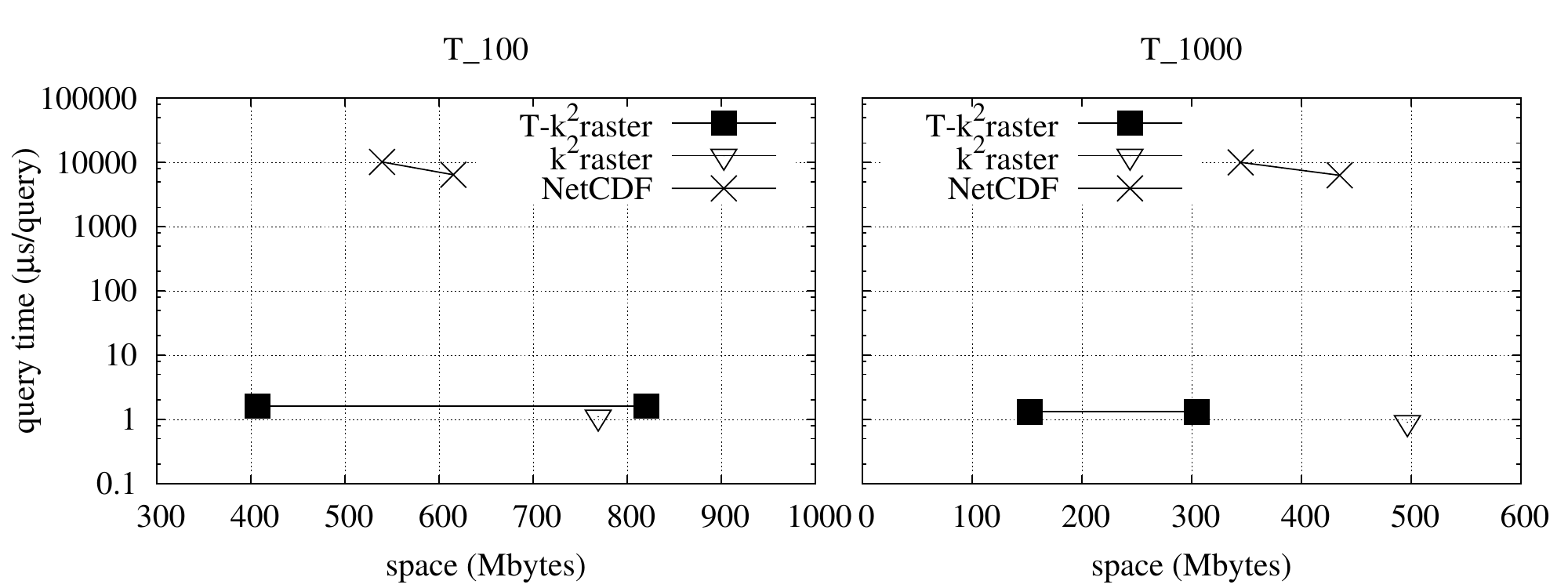}
  \vspace{-1cm}
  \begin{center}
  \caption{Space/time trade-off on T\_100 and T\_1000 datasets for cell value queries.}
  \label{fig:syntheticcelltimes}
  \end{center}
\vspace{-1cm}
\end{figure}

Figure~\ref{fig:syntheticcelltimes} shows the space/time trade-off for the datasets T\_100 and T\_1000 in cell value queries. We show the results only for NetCDF with compression enabled (deflate level 2 and 5), and for $\tkdosr$ with a sampling interval of 6 and 50. The $\tkdosr$ is slower than the baseline $\kdosr$, but is much smaller if a good $t_\delta$ is selected. Note that we use two extreme sampling intervals to show the consistency of query times, since in practice only the best approach in space would be used for a given dataset. In our experiments we work with a fixed $t_\delta$, but straightforward heuristics could be used to obtain an space-efficient $\tkdosr$ without probing for different periods: for instance, the number of nodes in the tree of differences and in the snapshot is known during construction, so a new snapshot can be built whenever the size of the tree of differences increases above a given threshold.

\begin{table}[th]
    \centering
    \setlength\tabcolsep{3pt}
     %\scalebox{0.95}{
    	{\scriptsize  
   \begin{tabular}{rc|rr|r|rr||rr|r|rr|}
   
         \cline{3-12}
   &  & \multicolumn{5}{|c||}{\texttt{T\_100}} & \multicolumn{5}{|c|}{\texttt{T\_1000}} \\
   \cline{3-12}
   &  & \multicolumn{2}{|c|}{\textbf{\tkdosr}} & \multirow{2}{*}{\textbf{\kdosr}} & \multicolumn{2}{|c||}{\textbf{\netcdf}} & \multicolumn{2}{|c|}{\textbf{\tkdosr}} & \multirow{2}{*}{\textbf{\kdosr}} & \multicolumn{2}{|c|}{\textbf{\netcdf}}\\
           \cline{1-2}
         \multicolumn{1}{|r}{\textbf{wnd}}   & \textbf{rng} & \multicolumn{1}{|c}{\textbf{6}} & \multicolumn{1}{c|}{\textbf{50}} & & \multicolumn{1}{|c}{\textbf{2}} & \multicolumn{1}{c||}{\textbf{5}} &  \multicolumn{1}{|c}{\textbf{6}} & \multicolumn{1}{c|}{\textbf{50}} & & \multicolumn{1}{|c}{\textbf{2}} & \multicolumn{1}{c|}{\textbf{5}}\\
         \hline
%\multicolumn{1}{|r}{\textbf{4}} & \multicolumn{1}{c|}{\textbf{1}} & 2.0 & 2.1 & 1.6 & 6050 & 9820 & 2.1 & 2.1 & 1.5 & 5930 & 9700 \\
%\multicolumn{1}{|r}{\phantom{x}} & \multicolumn{1}{c|}{\textbf{4}} & 2.3 & 2.5 & 1.8 & 6080 & 9830 & 2.0 & 2.1 & 1.5 & 5940 & 9760\\
%        \hline
\multicolumn{1}{|r}{\textbf{16}} & \multicolumn{1}{c|}{\textbf{1}} & 3.6 & 3.8 & 2.8 & 6130 & 10070 & 3.3 & 3.4 & 2.5 & 6160 & 10020\\
\multicolumn{1}{|r}{\phantom{x}} & \multicolumn{1}{c|}{\textbf{4}} & 5.1 & 5.5 & 3.6 & 6240 & 10100 & 3.5 & 3.5 & 2.6 & 6160 & 10100\\
        \hline
%\multicolumn{1}{|r}{\textbf{64}} & \multicolumn{1}{c|}{\textbf{1}} & 19.4 & 21.4 & 14.1 & 6760 & 10950 & 17.2 & 18.7 & 13.8 & 6730 & 10970\\
%\multicolumn{1}{|r}{\phantom{x}} & \multicolumn{1}{c|}{\textbf{4}} & 37.8 & 42.3 & 25.8 & 6820 & 10980 & 17.6 & 19.3 & 14.2 & 6650 & 10800\\
%        \hline
\multicolumn{1}{|r}{\textbf{256}} & \multicolumn{1}{c|}{\textbf{1}} & 222.9 & 248.1 & 163.9 & 9610 & 15330 & 207.1 & 228.9 & 167.6 & 9370 & 15110 \\
\multicolumn{1}{|r}{\phantom{x}} & \multicolumn{1}{c|}{\textbf{4}} & 429.3 & 489.4 & 301.7 & 9340 & 14790 & 213.4 & 234.3 & 172.7 & 9510 & 15240 \\
        \hline
\multicolumn{1}{|r}{\textbf{ALL}} & \multicolumn{1}{c|}{\textbf{1}} & 111450 & 126220 & 78250 & 443830 & 580660 & 79650 & 89380 & 63350 & 436400 & 568730\\
        \hline
\end{tabular}
}
%\normalsize
% }

    \caption{Range query times over \texttt{T\_100} and \texttt{T\_1000} datasets. Times shown in $\mu$s/query for different spatial windows (wnd) and range of values (rng).}\label{table:syntheticrangetimes}
\vspace{-6mm}
\end{table}

Table~\ref{table:syntheticrangetimes} shows an extract of the range query times for all the representations in datasets T\_100 and T\_1000. We only include here the results for $\tkdosr$ with a $t_\delta$ of 6 and 50, and for $\netcdf$ with deflate level 2 and 5, since query times with the other parameters report similar conclusions. We also show the results for some relevant spatial window sizes and ranges of values. In all the cases, $\tkdosr$ is around 50\% slower than $\kdosr$, due to the need of querying two trees to obtain the results. However, the much smaller space requirements of our representation compensate for this query time overhead, especially in T\_1000. $\netcdf$, that is not designed for this kind of queries, cannot take advantage of spatial windows or ranges of values, so it is several orders of magnitude slower than the other approaches. The last query set (\texttt{ALL}) involves retrieving all the cells in the raster that have a given value (i.e. the spatial window covers the complete raster). In this context, $\netcdf$ must traverse and decompress the whole raster, but our representation cannot take advantage of its spatial indexing capabilities, so this provides a fairer comparison. Nevertheless, both $\tkdosr$ and $\kdosr$ are still several times faster than $\netcdf$ in this case, and our proposal remains very close in query times to the $\kdosr$ baseline.

% \begin{figure}[htbp]
%   \vspace{-6mm}
%   \begin{center}
%   \includegraphics[width=0.49\textwidth]{figures/VAP.pdf}
%   \caption{Space/time tradeoff on VAP dataset for cell value queries.}
%   \label{fig:vapcelltimes}
%   \end{center}
%   \vspace{-1cm}
% \end{figure}
\begin{figure}
\begin{center}
\begin{minipage}{\textwidth}
	\begin{minipage}[m]{0.48\textwidth}   
    	\includegraphics[width=\textwidth]{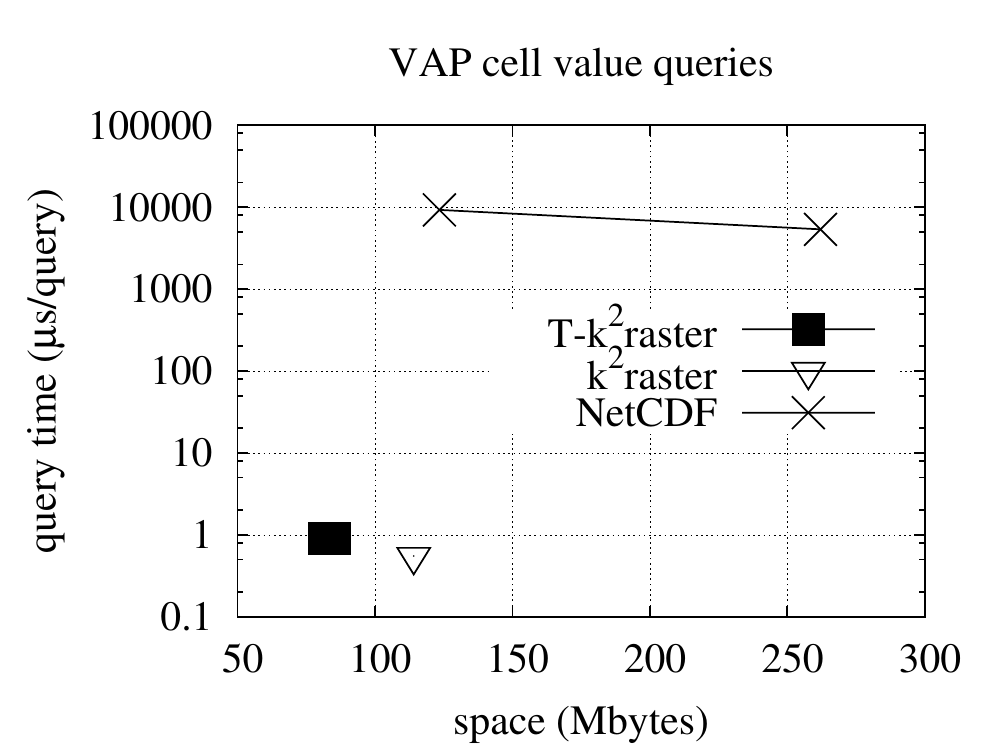}
%   		\captionof{figure}{Space/time tradeoff for cell value queries over VAP dataset.}
%   		\label{fig:vapcelltimes}
    \end{minipage}
    \begin{minipage}[m]{0.52\textwidth}
    \centering
    \setlength\tabcolsep{3pt}
     %\scalebox{0.95}{
	\scriptsize       
   \begin{tabular}{rc|rr|r|rr|}
         \cline{3-7}
            &  & \multicolumn{2}{|c|}{\textbf{\tkdosr}} & \multirow{2}{*}{\textbf{\kdosr}} & \multicolumn{2}{|c|}{\textbf{\netcdf}}\\
           \cline{1-2}
         \multicolumn{1}{|r}{\textbf{wnd}}   & \hspace{-1mm}\textbf{rng} & \multicolumn{1}{|c}{\textbf{6}} & \multicolumn{1}{c|}{\textbf{50}} & & \multicolumn{1}{|c}{\textbf{2}} & \multicolumn{1}{c|}{\textbf{5}}\\
         \hline
\multicolumn{1}{|r}{\textbf{4}} & \multicolumn{1}{c|}{\textbf{1}} & 2.2 & 2.2 & 1.6 & 5570 & 9520 \\
\multicolumn{1}{|r}{\phantom{x}} & \multicolumn{1}{c|}{\textbf{2}} & 2.0 & 2.0 & 1.4 & 5530 & 9430 \\
\multicolumn{1}{|r}{\phantom{x}} & \multicolumn{1}{c|}{\textbf{3}} & 1.9 & 1.8 & 1.3 & 5580 & 9430 \\
\multicolumn{1}{|r}{\phantom{x}} & \multicolumn{1}{c|}{\textbf{4}} & 1.7 & 1.7 & 1.3 & 5550 & 9470 \\
        \hline
\multicolumn{1}{|r}{\textbf{16}} & \multicolumn{1}{c|}{\textbf{1}} & 4.3 & 4.2 & 3.1 & 5670 & 9670 \\
\multicolumn{1}{|r}{\phantom{x}} & \multicolumn{1}{c|}{\textbf{2}} & 3.7 & 3.7 & 2.7 & 5630 & 9730 \\
\multicolumn{1}{|r}{\phantom{x}} & \multicolumn{1}{c|}{\textbf{3}} & 3.3 & 3.4 & 2.4 & 5660 & 9660 \\
\multicolumn{1}{|r}{\phantom{x}} & \multicolumn{1}{c|}{\textbf{4}} & 2.9 & 3.0 & 2.2 & 5720 & 9740 \\
        \hline
\multicolumn{1}{|r}{\textbf{64}} & \multicolumn{1}{c|}{\textbf{1}} & 26.4 & 26.1 & 19.5 & 6150 & 10470 \\
\multicolumn{1}{|r}{\phantom{x}} & \multicolumn{1}{c|}{\textbf{2}} & 21.1 & 21.6 & 16.1 & 6140 & 10440 \\
\multicolumn{1}{|r}{\phantom{x}} & \multicolumn{1}{c|}{\textbf{3}} & 16.8 & 17.3 & 12.9 & 6130 & 10450 \\
\multicolumn{1}{|r}{\phantom{x}} & \multicolumn{1}{c|}{\textbf{4}} & 16.2 & 16.6 & 12.6 & 6220 & 10660 \\
        \hline
\multicolumn{1}{|c}{\textbf{256}} & \multicolumn{1}{c|}{\textbf{1}} & 239.6 & 242.5 & 179.4 & 8720 & 14820 \\
\multicolumn{1}{|c}{\phantom{x}} & \multicolumn{1}{c|}{\textbf{2}} & 207.2 & 218.7 & 161.0 & 8660 & 14640 \\
\multicolumn{1}{|c}{\phantom{x}} & \multicolumn{1}{c|}{\textbf{3}} & 181.7 & 187.9 & 140.3 & 8590 & 14430 \\
\multicolumn{1}{|c}{\phantom{x}} & \multicolumn{1}{c|}{\textbf{4}} & 142.2 & 146.5 & 112.9 & 8300 & 14020 \\
        \hline
\multicolumn{1}{|c}{\textbf{ALL}} & \multicolumn{1}{c|}{\textbf{1}} & 60400 & 62900 & 46200 & 411700 & 552500\\
        \hline
\end{tabular}
%\normalsize
% }

%     \captionof{table}{Range query times (in $\mu$s/query) over \texttt{VAP} dataset.}\label{table:vaprangetimes}
    \end{minipage}
    {
   %\captionof{figure}{ Results for VAP dataset. Left plot shows space/time tradeoff for cell value queries. Right plot shows query times for range queries. Times in $\mu$s/query.}\label{fig:vapresults}
    }
\end{minipage}
\end{center}
\caption{ Results for VAP dataset. Left plot shows space/time tradeoff for cell value queries. Right table shows query times for range queries. Times in $\mu$s/query.}\label{fig:vapresults}
\end{figure}

Figure~\ref{fig:vapresults} (left) shows the space/time trade-off for the real dataset VAP. Results are similar to those obtained for the previous datasets: our representation, $\tkdosr$, is a bit slower in cell value queries than $\kdosr$, but also requires significantly less space. The $\netcdf$ baseline is much slower, even if it becomes competitive in space when deflate compression is applied.

Finally, Figure~\ref{fig:vapresults} (right) displays the query times for all the alternatives in range queries over the VAP dataset. The $\kdosr$ is again a bit faster than the $\tkdosr$, as expected, but the time overhead is within 50\%. $\netcdf$ is much slower, especially in queries involving small windows, as it has to traverse and decompress a large part of the dataset just to retrieve the values in the window. Note that even if the window covers the complete raster, $\tkdosr$ and $\kdosr$ achieve significantly better query times. 

%\newpage
%%%%%%%%%%%%%%%%%%%%%%%%%%%%%%%%%%%%%%%%%%%%%%%%%%%%%%%%%%%%%%%%%%%%%%%%%%%%%%%%
\section{Conclusions and future work} \label{sec:conclusions}
%%%%%%%%%%%%%%%%%%%%%%%%%%%%%%%%%%%%%%%%%%%%%%%%%%%%%%%%%%%%%%%%%%%%%%%%%%%%%%%%
In this work we introduce a new representation for time-evolving raster data. Our representation, called $\tkdosr$, is based on a compact data structure for raster data, the $\kdosr$, that we extend to efficiently manage time series. Our proposal takes advantage of similarities between consecutive snapshots in the series, so it is especially efficient in datasets where few changes occur between consecutive time instants. The $\tkdosr$ provides spatial and temporal indexing capabilities, and is also able to efficiently filter cells by value. Results show that, in datasets where the number of changes is relatively small, our representation can compress the raster and answer queries very efficiently. Even if its space efficiency depends on the dataset change rate, the $\tkdosr$ is a good alternative to compress raster data with high temporal resolution, or slowly-changing datasets, in small space. 

As future work, we plan to apply to our representation some improvements that have already been proposed for the $\kdosr$, such as the use of specific compression techniques in the last level of the tree. We also plan to develop an adaptive construction algorithm, that selects an optimal, or near-optimal, distribution of snapshots to maximize compression.

\end{document}